# Photonic Technologies for a Pupil Remapping Interferometer

Peter Tuthill[a], Nemanja Jovanovic[b,d], Sylvestre Lacour[c], Andrew Lehmann[b], Martin Ams[b,e], Graham Marshall[b,e], Jon Lawrence[b,d], Michael Withford[b,e], Gordon Robertson[a], Michael Ireland[a], Benjamin Pope[a], Paul Stewart[a]

[a]School of Physics, University of Sydney, NSW 2006, Australia;
[b]Macquarie Photonics Research Center, Dept. Physics and Astronomy, Macquarie University, NSW 2109 Australia;
[c]Observatoire de Paris, 5 place Jules Janssen, Meudon, France;
[d]Australian Astronomical Observatory, NSW 1710, Australia;
[e]Center for Ultrahigh bandwidth Devices for Optical Systems (CUDOS)

## ABSTRACT

Interest in pupil-remapping interferometry, in which a single telescope pupil is fragmented and recombined using fiber optic technologies, has been growing among a number of groups. As a logical extrapolation from several highly successful aperture masking programs underway worldwide, pupil remapping offers the advantage of spatial filtering (with single-mode fibers) and in principle can avoid the penalty of low throughput inherent to an aperture mask. However in practice, pupil remapping presents a number of difficult technological challenges including injection into the fibers, pathlength matching of the device, and stability and reproducibility of the results. Here we present new approaches based on recently-available photonic technologies in which coherent three-dimensional waveguide structures can be sculpted into bulk substrate. These advances allow us to miniaturize the photonic processing into a single, robust, thermally stable element; ideal for demanding observatory or spacecraft environments. Ultimately, a wide range of optical functionality could be routinely fabricated into such structures, including beam combiners and dispersive or wavelength selective elements, bringing us closer to the vision of an interferometer on a chip.

## 1. INTRODUCTION AND PRINCIPLES

The technique of aperture masking interferometry[1] lay as a dormant historical curiosity for most of the 20th century until its renaissance with John Baldwin's testbed demonstrating closure phase recovery at optical wavelengths with the William Herschel Telescope.[2] This experiment went on to produce pioneering work in the study of the surface structure and atmospheric stratification of red giants and supergiants.[3–7] A far wider realm of astrophysical targets was opened by the first major program on a 10 m class telescope: the Keck masking experiment.[8] Operating in the near-infrared and with snapshot two-dimensional Fourier mapping capability, this experiment revealed elaborate halos surrounding dying stars,[9–13] the first images of self-luminous disks circling young stars,[14–16] and the spectacular plumes around dusty Wolf-Rayets.[17–21]

The disproportionate success of such a simple, low-cost technique in the face of overwhelmingly greater investment in technological solutions such as Adaptive Optics provokes interesting questions. What are the underlying principles which allow masking to out-perform other methods in the recovery of high resolution structure at spatial scales in and around the formal diffraction limit? Furthermore, can these principles be extended and exploited to build still more advanced instruments for astronomical imaging?

The critical advantage offered by a pupil mask is that the telescope is transformed into a *non-redundant* array of sub-apertures. Redundant pupils will have two or more pairs of sub-apertures contributing to the same spatial frequency in the image plane. Under such conditions where the same baseline is so replicated, then in a seeing-limited case, the power will add incoherently. The resultant baseline power will then be a random walk of $R$ steps where $R$ is the *redundancy* – the number of times the given baseline is repeated within the pupil. In the bright source limit, the noise inherent to this random walk process completely dominates the signal-to-noise (SNR), leading to the well-known result that the SNR of speckle frames saturates to 1. For a non-redundant (masked) pupil, this random walk and the "seeing noise" which it precipitates is eliminated. Although a mask will discard most of the pupil area, and so lose a large fraction of the signal, it also removes almost all the atmospheric noise, and this leads to dramatically enhanced SNR.

Until recently, masking has been restricted to a niche for bright target astronomy due to the loss of light at the mask and the requirement for short (tens of milliseconds) exposure times. These problems have been largely overcome with the advent of "sparse aperture AO"[22] in which a masking experiment is operated in concert with an adaptive optics system. For a detailed discussion of this technique, in which the AO system acts as a "fringe tracker" permitting long exposures without the loss of all fringe visibility, see Tuthill et al., in this conference (7735-58). As described in this paper, this innovative synthesis of the two imaging methods has already opened rich new realms of astrophysical research in the detection of high contrast companions and in optical interferometric polarimetry.

Despite these promising advances, masking interferometry still suffers from several disadvantages. The great majority of the incident starlight (typically 90 – 98%) is still lost at the mask, and calibration is still affected by redundancy noise. To strictly enforce the condition of non-redundancy, a mask would require the idealization of infinitesimally small holes so that atmospheric noise is eliminated only when throughput goes to zero.

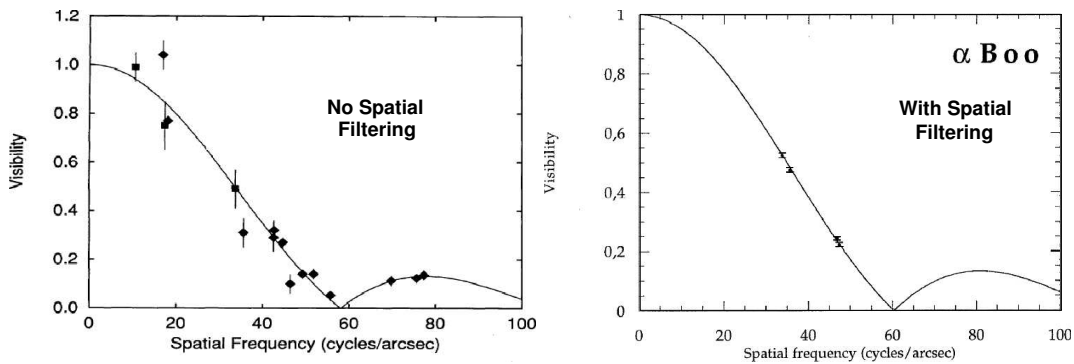

**Figure 1.** Visibility data recovered from optical long-baseline interferometers studying the red supergiant $\alpha$ Boo. The left-hand panel shows data points and best-fit model curve from the IRMA interferometer (no spatial filtering) while data from the FLOUR instrument at the IOTA array (with spatial filtering) are shown in the right-hand panel. Overwhelming signal-to-noise improvements are evidenced by (1) the scale of the error bars, and (2) the scatter of the points about the theoretical curve.

The rapid advance of fiber and optical waveguide technologies in recent years led several separate groups worldwide to independently propose a new imaging methodology: pupil remapping interferometry. Rather than use a mask to reject areas of the pupil, the entire pupil is fragmented into a grid with the use of a lenslet array which feeds each sub-pupil into a single-mode waveguide. This setup offers several critical immediate benefits;

- once light is in the waveguide, it is a straightforward matter to reconfigure the highly-redundant input pupil into a non-redundant output pupil simply by channeling the light

- single-mode waveguides act as spatial filters, rejecting all phase diversity across the input aperture so that the output beam will be a pure guided mode

- exquisite calibration can be achieved with spatial filtered beam recombination[23]

- these dramatic enhancements to data fidelity can be obtained with full pupil utilization and therefore significantly higher throughputs.

A graphic illustration which reinforces the point that spatial filtering can deliver an enormous boost to the signal-to-noise of visibility measurements in optical interferometry is given in Figure 1. Studying the same stellar target ($\alpha$ Boo) over the same region of spatial frequency, the figure contrasts the data fidelity from a free-space beamsplitter interferometer (IRMA) compared to the single-mode fiber spatial filter beam combiner of the FLUOR instrument.[24]

## 2. DESIGN OF A PUPIL REMAPPING INTERFEROMETER

Although a number of groups worldwide were active in experimentation with fiber optics in long baseline optical interferometry, the first published proposal for a single-telescope pupil remapping interferometer was the DAFI instrument concept[25–27] which appeared in the late 1990's. Several years later, a comprehensive series of papers exploring the technical and scientific aspects of the well-developed FIRST instrument concept[28, 29] from Meudon Observatory injected considerable excitement with quantitative projections of performance and signal-to-noise expectations for a working device.[30, 31] At around this time, the Sydney *Dragonfly* instrument obtained its first seed money, and together with one group in the US, the focus shifted from conceptual design studies to the difficult task of fabricating a working pupil remapping interferometer.

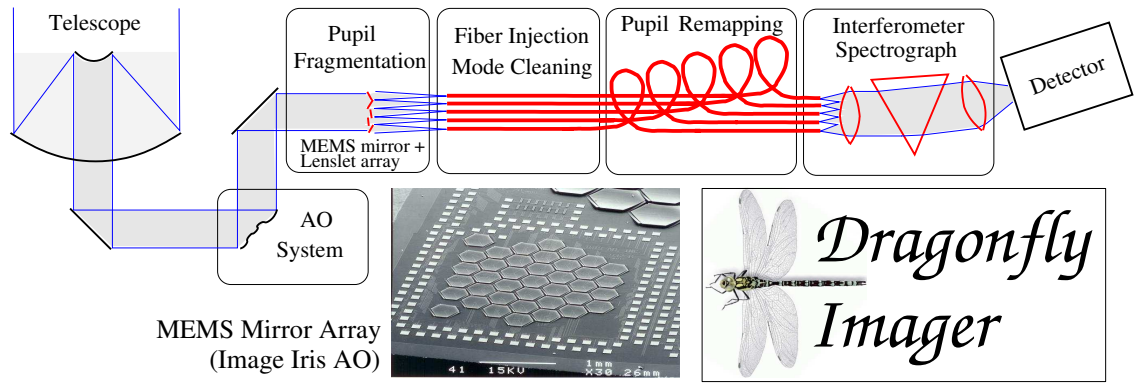

**Figure 2.** Schematic diagram showing the major subsystems of the *Dragonfly* concept. The telescope and AO system are assumed existent. Also shown is a scanning electron micrograph image of a working hexagonal MEMS mirror array consisting of 37 segments fabricated on a silicon substrate (Image courtesy of IRIS AO).

The initial conceptual design for Dragonfly is given in Figure 2. It is mounted within a telescope assumed to have an operational AO system that serves two distinct purposes: (1) an AO-stabilized wavefront has small tilt errors over individual sub-apertures and so can be injected into single-mode fibers with acceptable throughput, and (2) the phase of fringes formed between pairs of sub-apertures will be stabilized, allowing longer integration times. The remaining subsystems of Figure 2 are now discussed in turn.

A lenslet array placed in the beam will fragment the telescope pupil into sub-regions, which can then be injected into waveguides. In this case, these are polarization-preserving optical fibers arranged in a grid which precisely matches that of the lenslet array. At this stage, a MEMS micro-mirror array will be used to fine tune static wavefront errors which may affect coupling and coherence of the interferometer.

Once the light is injected into the single-mode fibers, spatial filtering is an automatic consequence, and pupil remapping is trivially enabled by plumbing the output fibers in a rearranged format to the input. For the prototype sketched in Figure 2, the remapped output pupil is a linear non-redundant array for Fizeau beam recombination. Fibers will be conveniently aligned in this format with precisely-fabricated silicon rulings, allowing injection directly into the interferometric spectrometer which cross-disperses the starlight into a number of separate wavelength channels. Spatial information will be conveniently encoded in one direction on the detector chip, with spectral dispersion on the orthogonal axis, in much the same manner as a number of interferometer-spectrographs such as the MIRC instrument[32] at CHARA.

# 3. TOWARDS AN INTEGRATED PHOTONIC INTERFEROMETER

The French "FIRST" and Australian *"Dragonfly"* instruments started with roughly similar conceptual designs, and as the practicalities of designing and sourcing optical elements progressed, the architectures began to converge (a process enhanced by travel and transfer of personnel between institutes). In order to rationalize these efforts and make progress in this technically demanding field without duplication, *Dragonfly* has now been redirected into a somewhat tangential development path (see below), while the original architecture depicted in Figure 2 is being actively pursued under the aegis of the FIRST project. This has reported very significant progress in recent years[33, 34] and has plans to go on-sky later this year.

At the heart of both *Dragonfly* and FIRST is the array of single-mode waveguides which perform the optical remapping of the pupil. The demands placed on the fabrication of this device are stringent: optical paths for all lightpaths must be matched to a tolerance that is a function of the single-channel bandwidth (there is also extra dependence associated with the way polarization is treated). These implementation-specific tolerances vary from the tens to hundreds of microns for most reasonable configurations, and they must be matched for the many fibers in a complex setup consisting of a precision fiber-array block at the input side and a silicon V-groove at the output. Practical experience from FIRST has shown that aligning and pathlength-matching the overall instrument poses very significant technical challenges. With meters of fiber-pathlength, care must be taken to guard against sensitivity to the thermal and mechanical environment at the telescope. Furthermore, an instrument completed along this conceptual design will be a hand-crafted triumph of specialized and high-precision metrology and micro-fabrication. The cycle time from prototype to working instrument is therefore lengthy and expensive, and will be compounded further if next-generation designs increase complexity with more waveguides or other functionality. For these reasons, *Dragonfly* is now exploring an alternative to the fiber-optic based architecture of Figure 2.

A very recent and exciting development in photonics is the use of powerful pulsed ultrafast-lasers to create high quality optical waveguides in many dielectric materials including both glasses and crystals.[35] By focusing femto- or picosecond pulses of sub-bandgap radiation into a bulk dielectric material, optical energy is deposited in the focal region in a multi-photon process. The deposited optical energy can induce a highly localized structural modification of the substrate material which, under the correct fabrication conditions, can result in a positive refractive index change. This refractive index modification can then be used to directly inscribe optical waveguides in a direct-write fashion by translating the material in three dimensions through the focus. The multi-photon nature of the process naturally delimits the optical microstructure to the immediate vicinity of the focal beam waist, and low-loss waveguides with similar properties to those obtained by drawing doped glasses or by photolithography have been demonstrated.

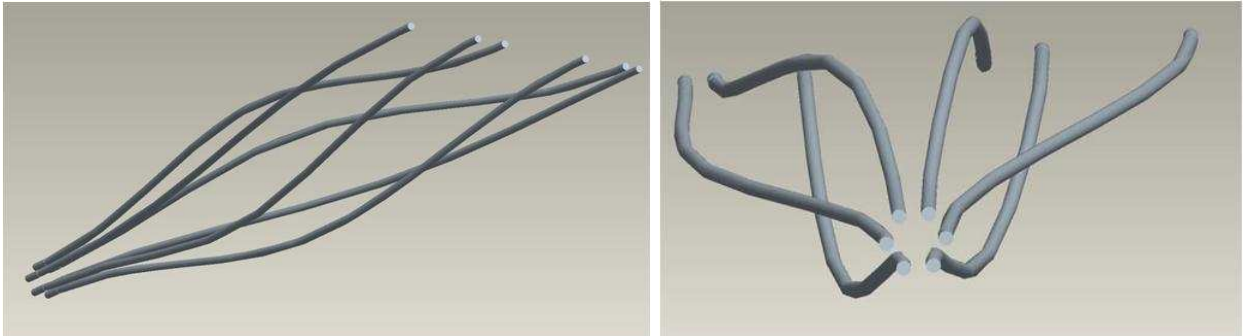

**Figure 3.** A conceptual 3-D sketch of single-mode waveguide tracks sculpted within a bulk substrate to perform the pupil remapping for the case of a simple six-element input array. Two different views onto the same structure are given in the panels above.

We propose to exploit this emergent technology to fabricate the most critical and technologically-challenging optical element which is central to our application: the pupil remapper. A concept drawing illustrating how this may work is given in Figure 3. Here we see at the input of the optical element a set of six waveguide tracks arranged in a regular hexagon which would receive light from a lenslet array of the same pitch. Light then follows the single-mode guides through the bulk substrate along trajectories which bring it to the output in a linear formation – in the case above this is a linear non-redundant array. Note that it would be equally easy to arrange the output guides with even spacings appropriate for butt-coupling to a photonic interferometer beam-combiner chip. In advance of manufacture, the design can be formulated so that there are nearly identical pathlengths along all tracks.

This fabrication method has immediate and profound benefits in the construction of a pupil remapping interferometer:

- Once a working design has been finalized, it is a relatively easy and mostly automated process to fabricate one or several components.

- A large number of complex optical tracks can be sculpted into a very small optical chip.

- Devices produced are robust, miniaturized, and are resiliant to thermal, mechanical and other environmental changes and stresses.

- Pathlength matching is expected to be relatively straightforward at the micron or sub-micron level without particular effort (this would present a very difficult task for a fiber-based device).

- Improvements between generations of prototype devices can be relatively easily achieved by modifying the computer-control fabrication template.

- Relatively short optical pathlengths make for better optical performance and less dispersion.

## 4. PROTOTYPING A PHOTONIC PUPIL REMAPPER

In order to test the viability of these photonic technologies in the fabrication of a working device, our first step was to construct very simple prototype devices. Our initial tests entailed two configurations sculpted into a single block with tracks designed to be single-moded in the optical at $\lambda = 0.78\mu$m where metrology, alignment and detection are straightforward with (nearly) visible light. Figure 4 shows a sketch of this first-generation test, which had one device (labelled the "tube") as a simple hexagonal array passing straight through the block, while the other (labelled the "fanout") was a transition between a hexagon at input and an evenly-spaced linear configuration at output. Although the tube structure is of no practical use in terms of pupil remapping, it does allow for the development of the protocol for the characterisation of such devices. Further, no care was taken in these devices to equalise the physical or optical path lengths of the fanout structure; instead smooth raised sine curves were used to route the waveguides from one end of the chip to the other.

Cosmetic investigation of the fabricated first-generation waveguides from images like those presented in Figure 4 revealed all structures were written to their design locations and there were no obvious defects within the glass. However, several guides did show evidence for ellipticity and other deformation from circularity of the cores (significantly worse than any depicted in Figure 4).

A program of detailed quantitative investigation of the waveguide properties was performed at the photonics laboratories in Macquarie University. Using a 780 nm laser, the near-field mode profiles of the waveguides were mapped, with the result that most tracks gave a nicely circular Gaussian-like mode profile like those depicted in Figure 5, although again there was a spread in performance with a few exhibiting notably elliptical mode patterns.

Although the results of the tests described so far are highly encouraging, further analysis did reveal some very significant defects in this first prototype device. A study was performed with a refractive index profilometer (Rinck Elektronik) which was able to measure the refractive index profile at 633 nm of the wavetracks within the substrate. These were found to have gaussian cross-sectional profiles (not a step-index) and were significantly distorted away from circular in many cases. However the worst aspect was a strong gradient in the core index step $\Delta n$ along the length of the device which started at values of $1.3 \times 10^{-3}$ at the LHS end through to about $3 \times 10^{-3}$ at the RHS end. This strong index taper was found to precipitate many further problems revealed by subsequent testing. The waveguides were found to be quite leaky – overall throughputs for the device were only around 10–20%. Flaring of the guided mode diameter due to the taper also promoted cross-talk between adjacent tracks, which were typically of order a percent, with a worst-case of 2.8%. Although these operational characteristics are significantly below expectations, the device has been extremely well characterized.

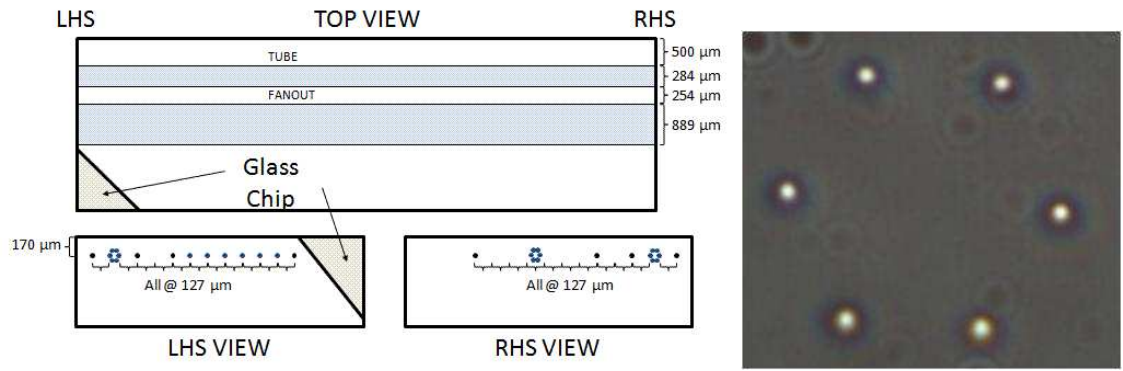

**Figure 4.** *Left panels:* a drawing of the 13 mm glass block into which our two trial geometries – the "Fanout" and the "Tube" – were carved by pulsed laser. Also shown are both left and right end-views of the block depicting the input geometry and output geometry of both configurations. *Right Panel:* A microscope image in white-light of the input array of the Fanout waveguide structure utilizing differential interference contrast to enhance the visibility of the structures.

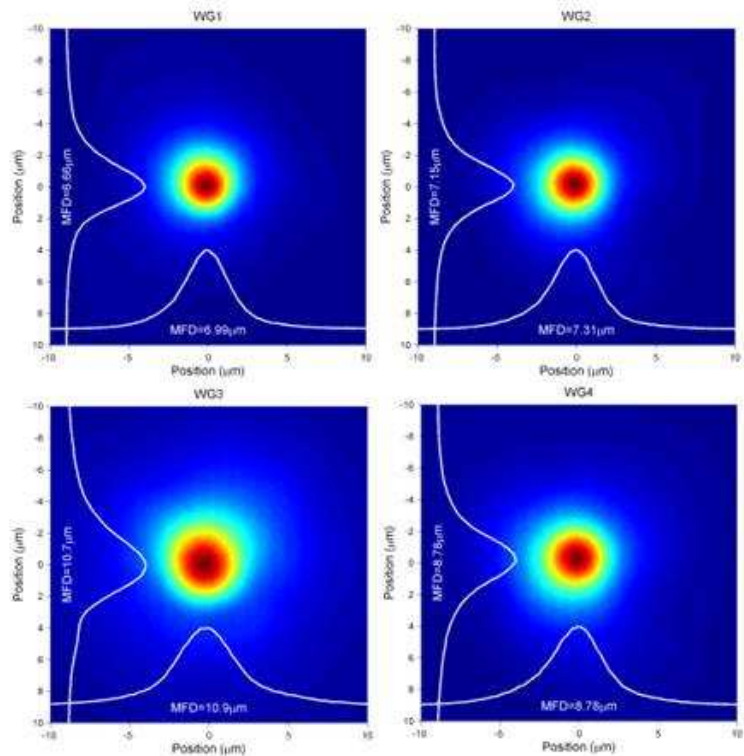

**Figure 5.** 780 nm flux profiles of four of the six waveguides from the hexagonal end of the "Fanout" device.

Numerical modelling with industry-standard optical waveguide design software (RSOFT) was performed with parameters reflecting our measurements: waveguide is a 3.5 μm Gaussian with $\Delta n = 3.9 \times 10^{-3}$ at one end with linear variation over 13 mm down to $\Delta n = 1.3 \times 10^{-3}$ at the other. The simulated injection was a SM600 fiber mode (4.3 micron core with $\Delta n = 5 \times 10^{-3}$). The model revealed that coupling is reasonable into the higher index end which guides well. However, as the light reaches the lower index regions of the taper, considerable leakage ensues. What is particularly encouraging is that the model has confirmed all performance metrics are to be expected from the structure as-written. Cross-coupling between guides was 2.5% and losses were around 80% – closely reflecting the measured values and implying that the underlying cause of the poor performance has been identified. A second model was generated with a constant $\Delta n = 3.9 \times 10^{-3}$ (removing the taper) and the unexpected problems with cross-coupling and leakage dissappeared.

It therefore appears that some unexplained drift or instability with the laser-write process has adversely compromised this first proof-of-concept device, and we are investigating several possibilities. Performance at these levels is not typical of work the system has delivered for other projects, and waveguides of high uniformity and with excellent refractive index profiles in the region of $\Delta n \sim 4 \times 10^{-3}$ have been routinely produced.

Despite these performance details, the exciting capability to sculpt a monolithic, miniaturized, single-mode pupil remapper has been clearly demonstrated. A working demonstrator could be constructed even with the imperfect device already fabricated, and there is every reason to believe that next-generation devices, recently completed (but not yet tested at the time of writing) will deliver levels of performance which more closely match expectations.

## 5. CONCLUSIONS AND FURTHER WORK

Our first-generation visible-light prototype has demonstrated the successful single-mode 2-D to 1-D array transition necessary for a successful pupil remapper. Some departures from expected performance levels were noted, but we believe that subsequent devices should perform more closely at required specifications. We have a second-generation of devices undergoing testing which are designed for the near-IR at 1550 nm and we will characterize these in terms of throughput, cross-talk, optical pathlength and polarization properties in the coming months. Our near-term goal is to try to build a prototype ready as an on-sky demonstrator. These devices can be used to feed into a (lithographically-fabricated) photonic chip beam combiner, or alternatively into a non-redundant output array configuration for recombination in a Fizeau image plane.

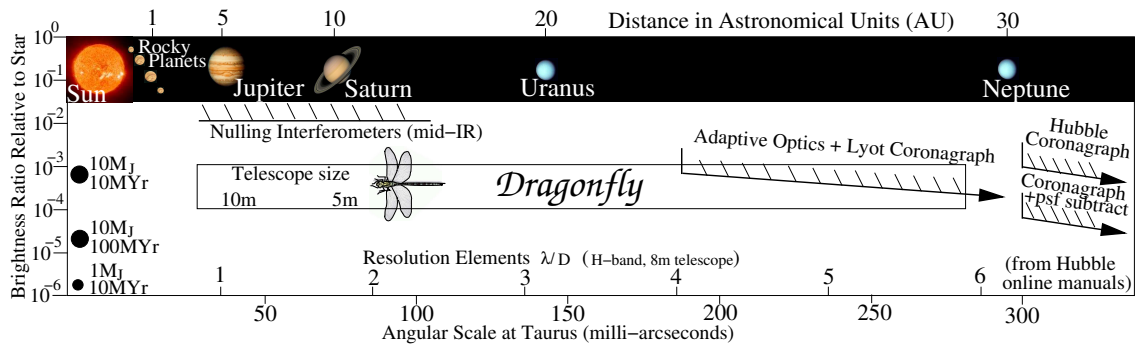

**Figure 6.** Diagram showing the discovery space explored by the *Dragonfly* instrument. Our own solar system is shown, with angular scale (lower axis) appropriate for 140 pc (Taurus distance), while hypothetical planets of varying Jupiter mass ($M_J$) & age are shown as a function of the dynamic range required on the vertical axis. The boxed region shows *Dragonfly's* target area, while hatched lines show existing ground/space coronagraphs and nulling interferometers.

Our longer-term goals for these technologies are to enhance the levels of precision calibration of the telescope point-spread function beyond those obtainable from the non-redundant aperture masking experiments described earlier. Based on conservative projections in improved calibration performance, the detection of high contrast companions should be enabled by this technology. In contrast to the most common Lyot coronagraphs which discard the inner $4\lambda/D$ of the image plane with the occulting spot, a pupil remapping interferometer will be most sensitive to planets interior to these scales. When considering the motivation to choose the best possible star-planet contrast, we are driven to the youngest hottest systems just formed, which in turn brings us to the nearest regions of active formation such as Taurus at 140 pc. When we project our own solar system to these distances, Figure 6 graphically illustrates the importance of the angular scales encompassed by the inner few $\lambda/D$: this is the place to look for analogs of gas giants Jupiter and Saturn.